\documentclass[aps,twocolumn,showpacs,prl,superscriptaddress,floatfix,longbibliography]{revtex4-1}

\usepackage{amsmath}
\usepackage{amssymb}
\usepackage{natbib}
\usepackage[breaklinks]{hyperref}
\usepackage{graphicx}
\usepackage[arrow, matrix, curve]{xy}
\usepackage{bm}
\usepackage{color}

\newcommand{\bra}[1]{\mbox{$\langle #1 |$}}
\newcommand{\bd}[1]{\boldsymbol{ #1 }}
\newcommand{\ket}[1]{\mbox{$| #1 \rangle$}}

\renewcommand{\H}{\mathcal{H}}
\DeclareMathAlphabet\mathbfcal{OMS}{cmsy}{b}{n}

\usepackage[usenames,dvipsnames]{xcolor}
\hypersetup{colorlinks=true, linkcolor=BrickRed, urlcolor=blue!50!black, citecolor=blue!50!black}

\DeclareTextFontCommand{\bbf}{\bfseries\em}

\usepackage[normalem]{ulem}

\makeatletter
\newcommand\hide@visible[1]{%
  \bgroup\fboxsep=.3ex\colorbox{Gray}{begin hide}%
  #1\colorbox{Gray}{end hide}\egroup%
}
\newcommand\hide@hidden[1]{%
  \bgroup\fboxsep=.3ex\colorbox{Gray}{hidden text}%
}
\newcommand\hide@invisible[1]{}
\newcommand\makevisible{\let\hide\hide@visible}
\newcommand\makehidden{\let\hide\hide@hidden}
\newcommand\makeinvisible{\let\hide\hide@invisible}
\makeatother
\makehidden

\begin{document}
\title{Diverging exchange force and form of the exact density matrix functional}

\author{Christian Schilling}
\email{christian.schilling@physics.ox.ac.uk}
\affiliation{Clarendon Laboratory, University of Oxford, Parks Road, Oxford OX1 3PU, United Kingdom}

\author{Rolf Schilling}
\email{rschill@uni-mainz.de}
\affiliation{Institut f\"ur Physik, Johannes Gutenberg-Universit\"at, D-55099 Mainz, Germany}

\date{\today}

\pacs{}

\begin{abstract}
For translationally invariant one-band lattice models, we exploit the \emph{ab initio} knowledge of the natural orbitals to simplify reduced density matrix functional theory (RDMFT). Striking underlying features are discovered: First, within each symmetry sector, the interaction functional $\mathcal{F}$ depends only on the natural occupation numbers $\bbf{n}$. The respective sets  $\mathcal{P}^1_N$ and $\mathcal{E}^1_N$ of pure and ensemble $N$-representable one-matrices coincide. Second, and most importantly, the exact functional  is strongly shaped by the geometry of the polytope $\mathcal{E}^1_N \equiv  \mathcal{P}^1_N $, described by linear constraints $D^{(j)}(\bbf{n})\geq 0$. For smaller systems, it follows as $\mathcal{F}[\bbf{n}]=\sum_{i,i'} \overline{V}_{i,i'} \sqrt{D^{(i)}(\bbf{n})D^{(i')}(\bbf{n})}$. This generalizes to systems of arbitrary size by replacing each $D^{(i)}$ by a linear combination of \{$D^{(j)}(\bbf{n})$\} and adding a non-analytical term  involving the interaction $\hat{V}$. Third, the gradient $\mathrm{d}\mathcal{F}/\mathrm{d}\bbf{n}$ is shown to diverge on the boundary $\partial\mathcal{E}^1_N$, suggesting that the fermionic exchange symmetry manifests itself within RDMFT in the form of an ``exchange force''. All findings hold for systems with non-fixed particle number as well and $\hat{V}$ can be \emph{any} $p$-particle interaction. As an illustration, we derive the \emph{exact} functional for the Hubbard square.
\end{abstract}

\maketitle

\paragraph{Introduction.---} Reduced density matrix functional theory (RDMFT) \cite{G75,C00,M07,PG16,SKB17} has the potential of overcoming the shortcomings and fundamental limitations of the widely used  density functional theory (DFT) \cite{HK64,PY95,GD95,J15}. Involving the full one-particle reduced density matrix (1RDM) $\gamma$ facilitates not only an exact description of the single particle potential energy, $\mathcal{U}[\gamma]\equiv \mbox{Tr}[\hat{U}\gamma]$, but also of the kinetic energy, $\mathcal{T}[\gamma]\equiv \mbox{Tr}[\hat{T}\gamma]$. It remains to derive accurate approximations to the interaction term $\mathcal{F}[\gamma]$. Moreover, RDMFT allows explicitly for fractional occupation numbers as it is required in the description of strongly correlated systems \cite{PG16}.
At the same time, involving the full 1RDM lies, however, also at the heart of possible disadvantages of RDMFT relative to DFT: While both methods avoid the use of exponentially complex $N$-electron wave functions, the 1RDM involves $d^2$ degrees of freedom compared to $d$ for the spatial density used in DFT, where $d$ is the basis set size. To be more specific, one often uses the spectral representation $\gamma \equiv \sum_{j}n_j |\varphi_j\rangle\!\langle \varphi_j|$ and then minimizes the total energy functional  $\mathcal{E}[\gamma]=\mathcal{T}[\gamma]+\mathcal{U}[\gamma]+\mathcal{F}[\gamma]$ with respect to the natural occupation numbers (NONs) $n_j$ \textit{and} natural orbitals $\ket{\varphi_j}$, separately. The dependence on the latter makes the minimization of $\mathcal{E}$ particularly difficult and one often encounters slow convergence (see, e.g., \cite{PE05}).

The general situation drastically changes in favour of RDMFT for the important class of periodic one-band lattice systems as studied in solid state physics. The 1RDM inherits the translational symmetry of the ground state \cite{D76} and the natural orbitals are known from the very beginning. They are given for all systems by plane waves (multiplied by some spin state). Thus, various possible disadvantages of RDMFT compared to DFT disappear and RDMFT simplifies \emph{de facto} to a NON-functional theory.

Based on this observation and the fact that in general the significance of symmetries in physics can hardly be overestimated,
we will explore in this letter the role of the translational symmetry within RDMFT and reveal universal and far-reaching consequences.
In that sense, our work complements previous studies of the homogeneous electron gas \cite{CP99,LHG07,SDLG08,LHZG09,LSDEMG09}, periodic polymers \cite{PO00,PO05} and of lattice systems \cite{SG95,C97,LSP00,HC01,LSP02,LSP04,SLMC05,TP11,SP11,TP12,CC13,TSP13,SP14,CFSB15,
SBRR15,KSPB16,SLP16,CMS16,MPR17,MPR17b,MTP18,MRP18} in which the crucial role of symmetries was not further explored. In particular, we determine the sets $\mathcal{P}^1_N$ and $\mathcal{E}^1_N$ of pure and ensemble $N$-representable 1RDMs and show that they coincide. Then, in the form of an analytic derivation, we discover the general \emph{form} of the exact functional $\mathcal{F}$ which will illustrate the fundamental role of one-body $N$-representability constraints. Finally, we show that the fermionic exchange symmetry manifests itself within RDMFT in the form of an ``exchange force'' which diverges on the boundary $\partial\mathcal{E}^1_N$ of the polytope $\mathcal{E}^1_N=\mathcal{P}^1_N$. All those universal features will be illustrated in two lattice cluster systems.

\paragraph{One-body $N$-representability constraints.---} We consider translationally invariant systems of $N$ electrons on a one-band lattice in $D$ dimensions with periodic boundary conditions and $L$ sites in each direction. Due to the translational invariance, the symmetry-adapted ``orbital'' part of the one-electron states are plane waves with momenta $\vec{k}=(2\pi/L)(\nu_1,\ldots, \nu_D)^t \equiv (2\pi/L)\vec{\nu}$, where $\nu_i = 0,1,\ldots, L-1$. The spin-orbitals follow as $|\vec{\nu}m\rangle$ ($m = \pm \frac{1}{2}$) and we introduce for the following the collective quantum number $q\equiv (\vec{\nu}m)$. On the $N$-fermion level, a symmetry-adapted basis is then given by the Slater determinants $\ket{\bd{q}}\equiv \ket{q_1,\ldots,q_N}$. The translational and spin symmetries decompose the $N$-fermion Hilbert space $\H$ into irreducible sectors $\H^{(Q)}$, $Q\equiv (\vec{K},M_z)$, each of which is spanned by the Slater determinants $\{\ket{\bd{q}}\}_{\bd{q} \in \mathcal{I}^{(Q)}}$ with total momentum $\vec{K}=\sum^N_{n=1} \vec{k}_n$ and magnetization $M_z=\sum^{N}_{n=1} m_n$. The respective set of configurations $\bd{q}$ is denoted by $\mathcal{I}^{(Q)}$.

The crucial observation is now that any two Slater determinants belonging to the same symmetry sector $Q$ differ in at least two entries $q_n$.  As a consequence the 1RDM $\bra{q}\gamma\ket{q'} =\mbox{Tr}[c_{q'}^\dagger c_q \hat{\Gamma}]$ for  an $N$-fermion density operator
$\hat{\Gamma}=\sum_{\bd{q},\bd{q'}\in \mathcal{I}^{(Q)}} \Gamma_{\bd{q}\bd{q'}} \ket{\bd{q}}\!\bra{\bd{q'}}$
(including pure states $\hat{\Gamma}\equiv \ket{\Psi}\!\bra{\Psi}$, $\ket{\Psi}=\sum_{\bd{q}\in \mathcal{I}^{(Q)}}\alpha_{\bd{q}}\ket{\bd{q}}$) is diagonal. Its diagonal elements, the NONs $\bd{n}=(n_q)$, are given by
\begin{equation}\label{NONvec}
\bd{n} = \sum_{\bd{q}\in \mathcal{I}^{(Q)}} \Gamma_{\bd{q}\bd{q}}\, \bd{v}_{\bd{q}}\,
\stackrel{\small{\hat{\Gamma}\equiv\ket{\Psi}\!\bra{\Psi}}}{=}\,\sum_{\bd{q}\in \mathcal{I}^{(Q)}} |\alpha_{\bd{q}}|^2\, \bd{v}_{\bd{q}}\,,
\end{equation}
where $\bd{v}_{\bd{q}}\equiv (\bra{\bd{q}}c_q^\dagger c_q\ket{\bd{q}})$ is the vector of spin-momentum occupation numbers of the Slater determinant state $\ket{\bd{q}}\!\bra{\bd{q}}$. Its entries are one whenever $q$ is contained in $\bd{q}$ and zero otherwise. Since any $\bd{n}$ is given as the convex combination of the vectors $\{\bd{v}_{\bd{q}}\}_{\bd{q}\in \mathcal{I}^{(Q)}}$, the respective sets $\mathcal{E}_N^1(Q)$ and $\mathcal{P}_N^1(Q)$ of ensemble and pure $N$-representable 1RDMs are given as the polytope with vertices $\{\bd{v}_{\bd{q}}\}_{\bd{q}\in \mathcal{I}^{(Q)}}$ and in particular they do coincide (cf. Eq.~(\ref{NONvec})),
\begin{equation}\label{PeqE}
\mathcal{P}_N^1(Q) = \mathcal{E}_N^1(Q) \,.
\end{equation}
Since not all vertices of the hypercube $[0,1]^d$ with particle number $N$ contribute to those sets, the $N$-representability constraints for each sector $Q\equiv (\vec{K},M_z)$ are more restrictive than Pauli's exclusion principle \mbox{$0\leq n_q\leq 1$}. Yet, it is important to notice that the calculation of those symmetry-adapted generalized Pauli constraints is considerably simpler than the calculation of the one-body pure $N$-representability constraints for systems without symmetries.

As an illustration, we consider  three fully polarized electrons on a ring of six lattice sites with \mbox{$K=0$} (for details, see supporting information \footnote{See the Supplemental Material at url for technical details on the derivation of the exact functional, the diverging exchange force and the solution of lattice cluster systems, which includes Refs.~\cite{K09,SBV17,BD72,R07,F91}}). It is an elementary exercise to determine all $(\nu_1,\nu_2,\nu_3)$ with $\sum_{n=1}^3 \nu_n \,(\mbox{mod}\,6) =0$. One gets $(0,1,5)$, $(0,2,4), (1,2,3), (3,4,5)$ and the respective polytope \eqref{PeqE} is then given by the convex hull of the four vertices $(1,1,0,0,0,1), (1,0,1,0,1,0),(0,1,1,1,0,0)$ and $(0,0,0,1,1,1)$. By solving linear equations this vertex representation of  $\mathcal{P}_N^1 = \mathcal{E}_N^1$ can be transformed into a half space representation, $\{D^{(j)}(\bd{n}) \geq 0\}$, with the following four $N$-representability constraints:
\begin{eqnarray}\label{set36}
D^{(1)}({\bbf{n}})&=& n_0+n_1-n_2 \geq 0 \nonumber \\
D^{(2)}({\bbf{n}})&=& n_0-n_1+n_2  \geq 0\nonumber\\
D^{(3)}({\bbf{n}})&=& 2 - n_0-n_1- n_2 \geq 0\nonumber \\
D^{(4)}({\bbf{n}})&=& -n_0+n_1+n_2 \geq 0\,,
\end{eqnarray}
with the linearly dependent variables $n_3=1-n_0$, $n_4=1-n_1$ and $n_5=1-n_2$.
For larger settings, the easy-to-determine vertex representation of \eqref{PeqE} can be transformed into a half space representation by resorting to standard softwares.

\paragraph{Interaction functional $\mathcal{F}$ and exchange force.---} To elaborate on the structure of the \emph{exact} interaction functional $\mathcal{F}$, we resort to Levy's construction \cite{LE79} (see also Ref.~\cite{LI83}). For general systems (and by ignoring possible symmetries), the \textit{exact} $\mathcal{F}[\gamma]$ follows as the minimization of the interaction energy over the set of all $N$-fermion \textit{pure} states $\ket{\Psi}$ with 1RDM $\gamma\in \mathcal{P}_N^1$, i.e.~$\mathcal{F}_p[\gamma]= \min_{\Psi \mapsto \gamma} \langle \Psi |\hat{V}| \Psi \rangle$.
This leads to a ``pure RDMFT'' on $\mathcal{P}_N^1$. In practice, one tries, however, to avoid the
highly intricate generalized Pauli constraints \cite{BD72,KL06,AK08} by relaxing the minimization to $N$-fermion \emph{ensemble} states  $\hat{\Gamma}$ \cite{V80}. This then leads to an ``ensemble RDMFT'' with an interaction functional $\mathcal{F}_e$ on the set $\mathcal{E}^1_N$ which is described by the simple Pauli exclusion principle constraints only \cite{C63}.
Yet, this cannot allow one to ``circumvent'' the mathematically proven complexity of the ground state problem \cite{LCV07,SV09} and the complexity is just shifted from the set of underlying 1RDMs to the derivation of the functional $\mathcal{F}_e$ and/or its minimization \cite{S18}.
In that context, with regard to approximated functionals such as \cite{CP99,LHG07,SDLG08,LHZG09,LSDEMG09,M84,GU98,CA00,B01,Y01,Y02,BB02,CGA02,HH03,CBZ03,KH03,KH04,PC04,K04,
GPB05,P05,K06,RPGB08,LM08,ML08,PLRMU11,BV12,P12,PE13,PML13,PU14,P17,SKB17,P18,MRP18,BM18}, it is unclear why those
based on pure state ansatzes with fixed $N$ are treated within ``ensemble RDMFT'', as well. For more details the reader is
referred to the reviews \cite{PG16,SKB17} and references therein.

As already stressed above, for periodic one-band lattice systems the interaction functionals simplify drastically to functionals (or more precisely to functions) of the spin-momentum occupation numbers $\bd{n}$. For each $Q\equiv (\vec{K},M_z)$, Levy's construction \cite{LE79} is restricted to $\ket{\Psi}$ in the respective symmetry-sector (see also Refs.~\cite{G18, WK18})
\begin{equation} \label{LevyQ}
\mathcal{F}_p[\bbf{n}]= \min_{\H^{(Q)}\ni\Psi  \mapsto \bbf{n}} \langle \Psi |\hat{V}| \Psi \rangle \,.
\end{equation}
In the following, we simplify the notation by enumerating all configurations $\bd{q} \in \mathcal{I}^{(Q)}$, denote the respective Slater determinants by $\ket{r}$, $r=1,\ldots, R\equiv \dim{(\H^{(Q)})}$ and introduce $V_{r r'}\equiv \bra{r}\hat{V}\ket{r'}$. Moreover, we will focus on $\mathcal{F}_p$. As it is proven in the supporting information \cite{Note1}, the equivalence $\mathcal{F}_e\equiv \mathcal{F}_p$ holds, at least whenever there exists phase factors $\eta_r$ such that $V_{rr'}\equiv-\eta_r \eta_{r'}|V_{r r'}|$.

It is instructive to derive in a first step our main results for systems  in which $\mathcal{P}_N^1$ takes the form of a simplex, i.e., each of its facets contains all vertices except one. Equivalently, it means that the number of independent coefficients, $\{\alpha_{\bd{q}}\}$, equals the number of independent NONs, ${\bf{n}}$. This condition is valid for several smaller systems, but also for systems of arbitrary size in case their underlying Hilbert space is restricted within \eqref{LevyQ} to a subspace involving only $\mathcal{O}(d)$ CI coefficients  (yielding an approximate functional).
A prime example is the one of three fully polarized electrons on six sites as already discussed above (for details see \cite{Note1}).  We thus label the one-body $N$-representability constraints $D^{(r)}(\bbf{n})\geq 0$ such that the respective facet does not contain the vertex $\bd{v}_r$, i.e.~we have $D^{(r)}(\bd{v}_{r'})=0$ whenever $r\neq r'$. Moreover, we ``normalize'' each $D^{(r)}\geq 0$ such that $D^{(r)}(\bd{v}_r) =1$. Using Eq.~\eqref{NONvec} and the linearity of $D^{(r)}$, we find
\begin{equation}\label{DvsAsimplex}
D^{(r)}(\bbf{n}) = |\alpha_r|^2\, .
\end{equation}
It is exactly the simplicial structure of $\mathcal{P}_N^1$ which implies this crucial one-to-one relation between $\{D^{(r)}(\bbf{n})\}$ and $\{|\alpha_r|^2\}$. Consequently, Levy's construction \eqref{LevyQ} with the ansatz $\ket{\Psi}=\sum_{r} \eta_r |\alpha_r|\, \ket{r}$ is trivial to carry out up to the phase factors $\eta_r$ of $\alpha_r$. Their minimization leads to some $\overline{\eta}_r\equiv \overline{\eta}_r(\bbf{n},\hat{V})$ and eventually we obtain
\begin{equation}\label{Fsimplex}
\mathcal{F}_p[{\bbf{n}}]= \sum_{r,r'}  V_{rr'}\overline{\eta}_r^\ast\, \overline{\eta}_{r'} \sqrt{D^{(r)} ({\bbf{n}})\, D^{(r')} ({\bbf{n}})} \,.
\end{equation}
The result \eqref{Fsimplex} for the exact interaction functional valid for \emph{any} symmetry-respecting interaction $\hat{V}$ could hardly be more striking: $\mathcal{F}_p$ is fully determined (up to phase factors $\overline{\eta}_r(\bbf{n},\hat{V})$) by the geometry of the simplex $\mathcal{P}_N^1$. Moreover, the presence of an exchange force, as we shall call it, follows immediately which diverges on the boundary of $\mathcal{P}_N^1$,
\begin{equation}\label{ExFsimplex}
\left|\frac{\mathrm{d}\mathcal{F}_p}{\mathrm{d}\bbf{n}}[{\bbf{n}}]\right| \sim \mathcal{G}^{(r)}\frac{1}{\sqrt{D^{(r)}({\bbf{n}})}} \,,\quad\mbox{as}\,\,D^{(r)}({\bbf{n}})\rightarrow 0\,.
\end{equation}
Remarkably, the exchange force is always repulsive in the sense that it is repelling $\bbf{n}$ from the polytope boundary (see supporting information \cite{Note1}).

Generalizing the results \eqref{Fsimplex} and \eqref{ExFsimplex} to systems with \emph{arbitrary} underlying polytope $\mathcal{P}_N^1\equiv \mathcal{E}_N^1$ is quite intricate: Relation \eqref{DvsAsimplex} takes the form (see supporting information \cite{Note1})
\begin{equation}\label{DvsA}
D^{(j)}(\bbf{n}) = \sum_{r=1}^R D^{(j)}(\bd{v}_r)\, |\alpha_r|^2\,,
\end{equation}
for all $j=1,\ldots,J$, where typically $D^{(j)}(\bd{v}_r) > 0$ for more than one $r$. We also introduced $J$, the number of $N$-representability constraints. As a consequence, $\bbf{n}$ does not uniquely determine $\{|\alpha_r|\}$ anymore and instead a set of $d$ linear equations with $R>d$ variables has to be solved. The constrained search in \eqref{LevyQ} then amounts to a non-trivial minimization over the $R-d$ remaining variables.
This purely technical and less informative derivation (see supporting information \cite{Note1}) leads to the general final \textit{form}
\begin{eqnarray}\label{F}
&&\mathcal{F}_p[{\bbf{n}}]  =  \! \sum_{r,r'=1}^R \!V_{rr'}\, \overline{\eta}_r^\ast \,\overline{\eta}_{r'} \sqrt{\tilde{D}_r({\bbf{n}}, \hat{V})} \,\sqrt{\tilde{D}_{r'}({\bbf{n}}, \hat{V})}\,, \\
&&\tilde{D}_r({\bbf{n}}, \hat{V}) \equiv \sum^{J}_{j=1}b^{(j)}_r D^{(j)}({\bbf{n}})+\overline{a}_r\big(\{D^{(i)}({\bbf{n}})\}, \hat{V}\big) \nonumber \,.
\end{eqnarray}
The coefficients $b^{(j)}_r$ are solely determined by the geometry of the polytope $\mathcal{P}_1^N$ and $\overline{a}_r(\{D^{(j)}({\bbf{n}})\}, \hat{V})$ follow from the minimization of the degrees of freedom not fixed by $\bd{n}$. This highly involved minimization, as discussed in the supporting information \cite{Note1}, leads to an implicit additional dependence of $\mathcal{F}_p$ on ${\bbf{n}}$ and the interaction $\hat{V}$.

At the same time, the general form \eqref{F} offers excellent prospects for a perturbation theoretical approach by expanding $\overline{a}_r(\{D^{(j)}({\bbf{n}})\}, \hat{V})$ (see Hubbard square below).

Whenever ${\bbf{n}}$ approaches the facet described by $D^{(j)}\equiv 0$, it follows from Eq.~(\ref{DvsA}) that $|\alpha_r| \to 0$  for all $r$ whose vertices ${\bbf{v}}^{(r)}$ do not belong to that facet. This fact must reflect itself in the ${\bbf{n}}$-dependence of
$\mathcal{F}_p$. Indeed, one obtains for each $j$ the singular ${\bbf{n}}$-dependence \cite{Note1}
\begin{equation}\label{ExF}
\mathcal{F}_p [{\bbf{n}}] = \mathcal{F}_p^{(j)} + \mathcal{G}^{(j)}_p \sqrt{D^{(j)}({\bbf{n}})}
 + \mathcal{O}(D^{(j)}({\bbf{n}})) \,.
\end{equation}
This result presents in a particularly striking form the crucial role of the $N$-representability constraints $D^{(j)}({\bbf{n}})\geq 0$. In particular, as an extension of \eqref{ExFsimplex}, it confirms that the fermionic exchange symmetry manifests itself within RDMFT in the form of an exchange force diverging on the boundary of the polytope $\mathcal{P}_N^1=\mathcal{E}_N^1$.

\paragraph{Hubbard square.---} Now, as an illustration, we apply the general framework from above to the one-dimensional one-band Hubbard model with $N=4$ electrons, $L=4$ sites (half filling) and nearest neighbor hopping with hopping rate $t>0$. This will emphasize from a different perspective the drastic simplification of RDMFT in case all symmetries are fully exploited: The boundaries of exact functional calculation are extended from the commonly studied Hubbard dimer \cite{LSP00,LSP02,LSP04,CFSB15,SBRR15,KSPB16,CMS16} with an underlying six-dimensional Hilbert space to the Hubbard square with a Hilbert space of dimension $70 =\binom{8}{4}$.

The kinetic energy functional for the Hubbard square reads $\mathcal{T}[{\bbf{n}}]=-4t\sum_{\nu=0}^3 \cos{(2 \pi {\nu}/4)} (n_{\nu\uparrow}+n_{\nu\downarrow})$ and the Hubbard on-site interaction has strength $U\geq 0$ (Coulombic repulsion). We will present only the essential steps and refer to the supporting information \cite{Note1}, where all details of the following discussion are presented.
\begin{figure}[htb]
\includegraphics[width=0.67\columnwidth,height=4.0cm]{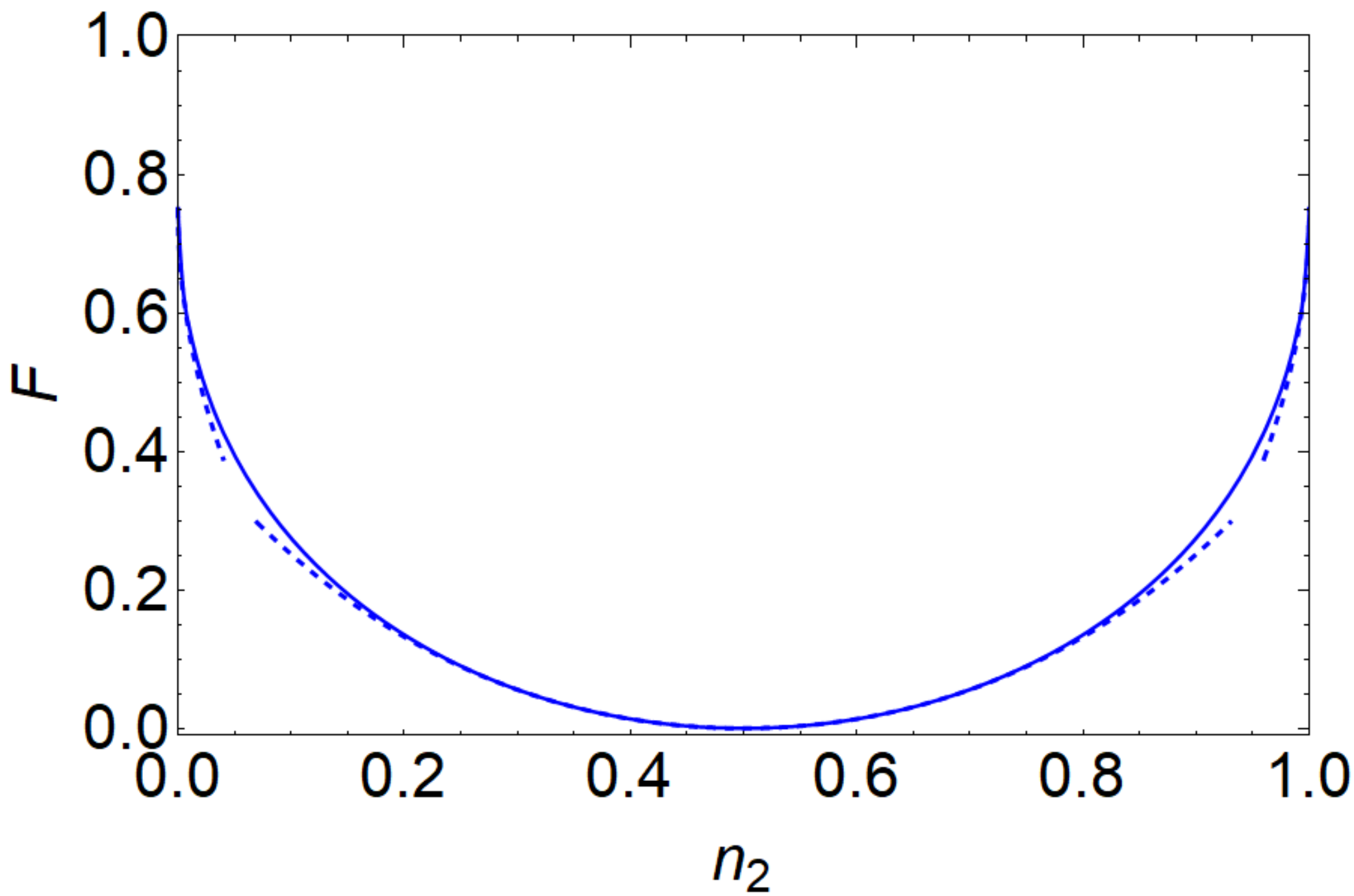}
\caption{Weak and strong coupling asymptotes $\eqref{eq13}$ (dashed lines) and exact functional $\mathcal{F}$ (solid line).}
\label{fig:functional}
\end{figure}

The ground state for  $U \geq 0$ is a singlet state with  total momentum  $K=\frac{2\pi}{4}2=\pi$  and parity $p=-1$. Taking all these symmetries into account leads to a rather simple polytope $\mathcal{P}_N^1=\mathcal{E}_N^1\cong[0,1]$ of $N$-representable 1RDMs: It is $n_{\nu\uparrow}=n_{\nu\downarrow}\equiv n_{\nu}$, $n_1=n_3=1/2$ and $n_0=1- n_2$. Hence there is  only one independent variable ($n_2$) (which is identified with ${\bbf{n}}$)  constrained by Pauli's exclusion principle $0 \leq n_2 \leq 1$, only. This is a particular \emph{incidence} and in larger systems in a singlet state, the translational symmetry  implies constraints which are more restrictive than Pauli's exclusion principle.

For given ${\bbf{n}}$, Levy's construction \eqref{LevyQ} cannot be fully carried out by analytical means since it involves the root of a polynomial of degree six. The exact functional $\mathcal{F}\equiv \mathcal{F}_p=\mathcal{F}_e$ \cite{Note1} as function of $n_2$ is determined numerically instead and we depict it in Figure 1. Its graph demonstrates the divergence of the slope on the ``facets'' $n_2=0,1$, as predicted by \eqref{ExF}. Also the particle-hole duality $\mathcal{F}[n_2] = \mathcal{F}[1-n_2]$  \cite {Y01} is obvious and the convexity of $\mathcal{F}$ is consistent with the fact that ``ensemble functionals'' $\mathcal{F}_e$ are always convex \cite{LI83,ZM85}.

Using a \emph{perturbative approach} for \eqref{F}, the functional $\mathcal{F}$ simplifies in the asymptotic regimes of weak ($0 \leq U \ll t$) and strong ($U \gg t$) coupling \cite{Note1},
\begin{eqnarray}\label{eq13}
\mathcal{F}[{\bbf{n}}]&=&  U\Big[\frac{3}{4} - \frac{\sqrt{13}}{2} \sqrt{n_2} + \mathcal{O}(n_2)\Big],  \, 0\leq U \ll t  \\
\mathcal{F}[{\bbf{n}}]&=&U\Big[\frac{4}{3}\big(\frac{1}{2}-n_2\big)^2  + \frac{40}{27}\big(\frac{1}{2}-n_2\big)^4 + \ldots\Big], \,  U \gg t\,.
\nonumber
\end{eqnarray}
\begin{figure}[htb]
\includegraphics[width=4.10cm]{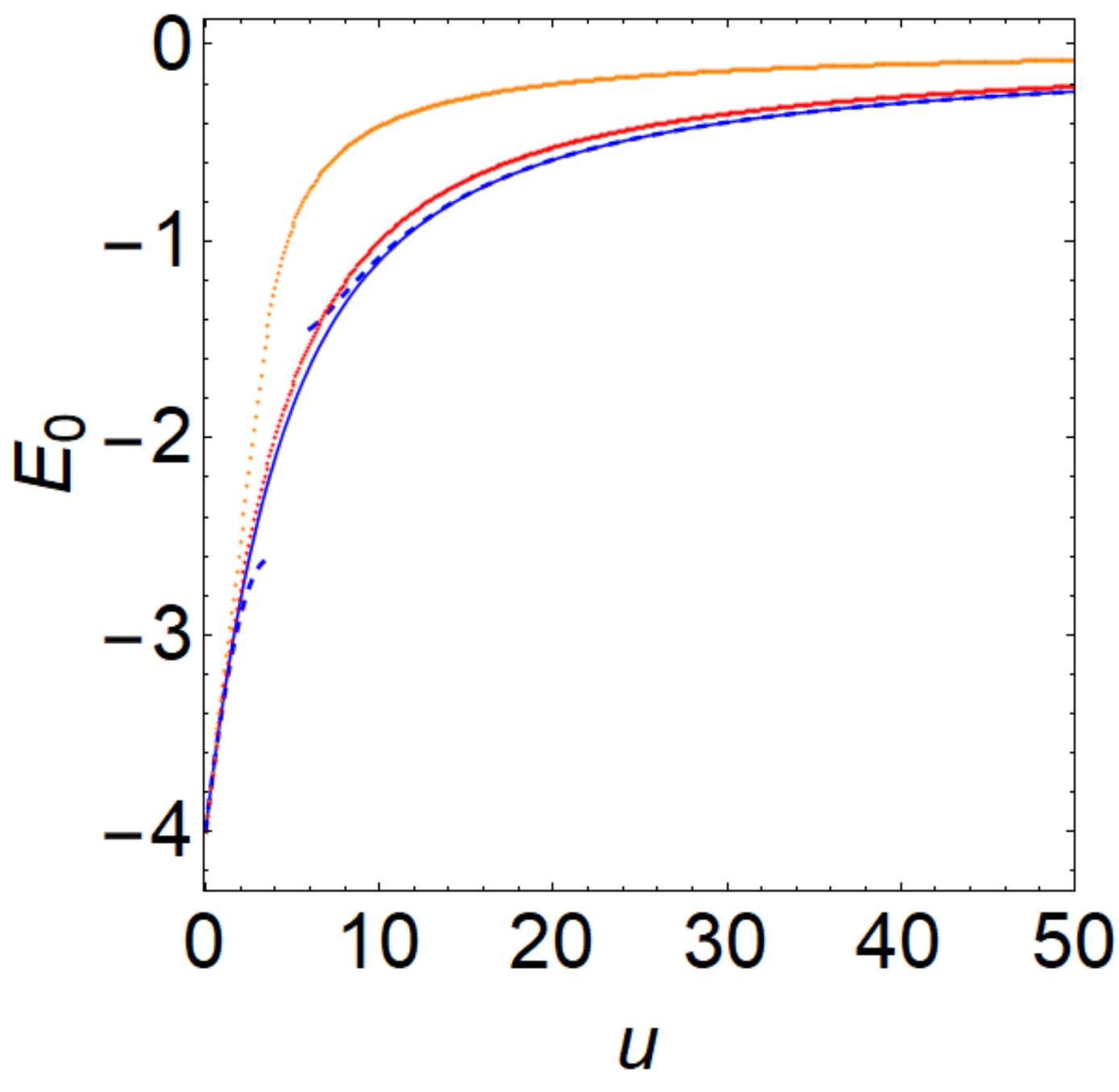}
\hspace{0.1cm}
\includegraphics[width=4.16cm]{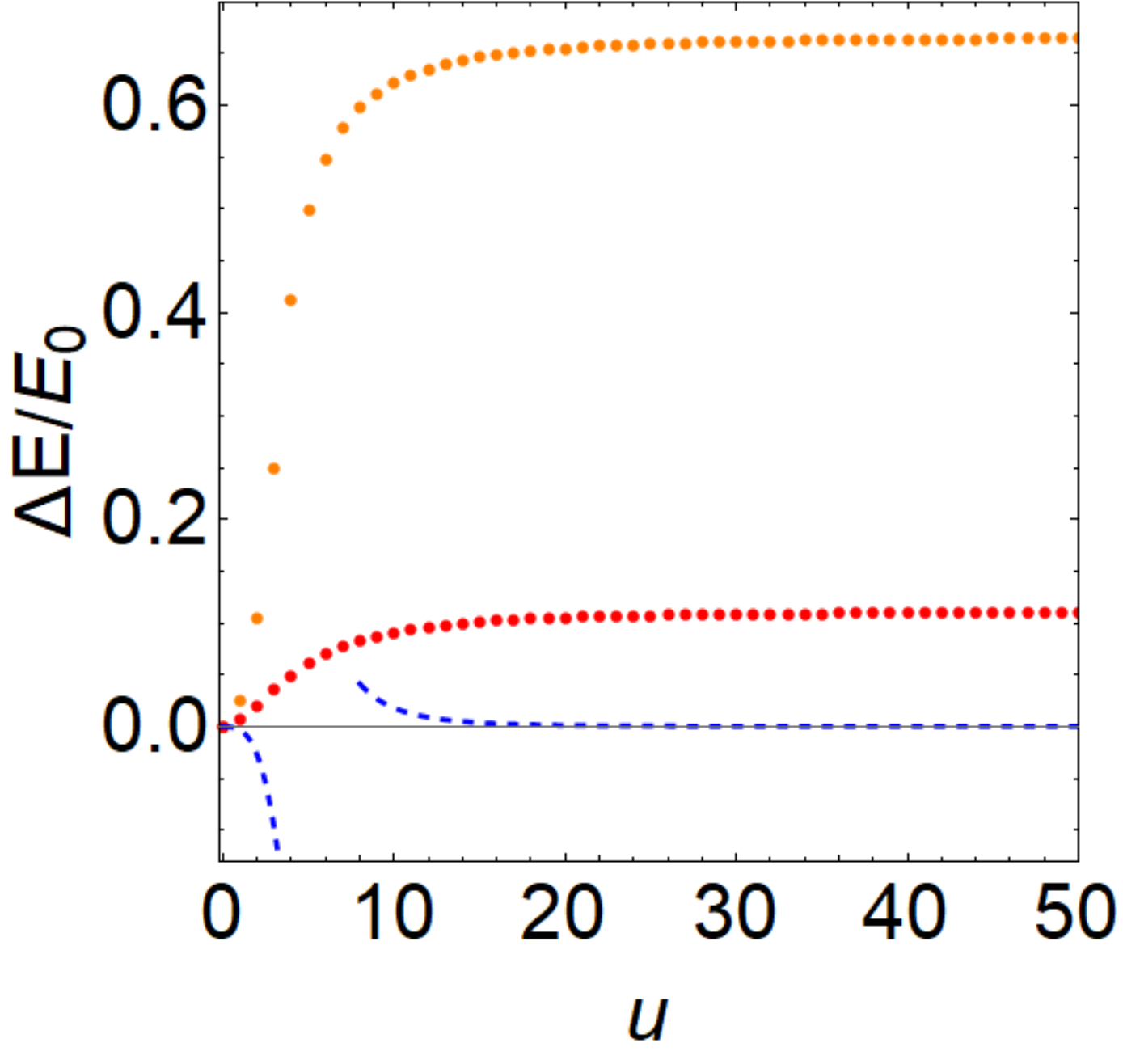}
\caption{Left: Exact result for the ground state energy  $E_0(u)$(blue solid line) from the exact functional. The weak and strong coupling result from the functionals (\ref{eq13}) is shown by the blue dashed lines. The result from PNOF5 and PNOF7(-) is presented by orange and red dots, respectively. Right: Relative error  $\Delta E/E_0$ as a function of $u$.}
\label{fig:E_0}
\end{figure}

Using $\mathcal{T}[{\bbf{n}}]= -8t(\frac{1}{2}-n_2)$ and the
results from Eq.~(\ref{eq13}), one obtains from the minimization of  $\mathcal{E}[{\bbf{n}}]$ the ground state energy $E_0$ and the corresponding NON $n_2$ in the weak coupling regime as a function of $u=U/t$
\begin{eqnarray}\label{eq14}
 E_0(u)/t &=&  -4 + \frac{3}{4}u - \frac{13}{128}u^2 + \mathcal{O}(u^{3}) \nonumber \\
 n_2(u) &= & \frac{13}{1024}u^2 + \mathcal{O}(u^{3})
\end{eqnarray}
and for strong coupling
\begin{eqnarray}\label{eq15}
E_0(u)/t &=& -12u^{-1} + 120 u^{-3} + \mathcal{O}(u^{-5}) \nonumber \\
n_2(u) &= &\frac{1}{2}- 3u^{-1} - 60 u^{-3}+ \mathcal{O}(u^{-5})\, .
\end{eqnarray}
The asymptotically exact results (\ref{eq14}),(\ref{eq15}) are shown in Figure 2 (left).
This figure also contains the exact result and those of PNOF5 \cite{PLRMU11,PML13} and PNOF7(-) \cite{MRP18}, the best approximate functionals among all used in Ref.~\cite{MPR17,MPR17b}. Result (\ref{eq15}) fits perfectly  the exact result for all $u > 10$. The convergence to zero for $u \to \infty$ (a general property of the Hubbard model at half filling in any dimension \cite{F91}) is reproduced also by PNOF5 and PNOF7(-). In order to check the quality of the approximate functionals more, we have also plotted the relative error $\Delta E/E_0$ in Figure 2 (right). We observe  that this error  is about $60\%$ and $10\%$ for PNOF5 and PNOF7(-), respectively, and practically zero for our approximate result (\ref{eq15}) for all $u > 10$.

\paragraph{Summary and conclusions.---} We have demonstrated how the \emph{ab initio} knowledge of the natural orbitals
for translationally invariant one-band lattice models significantly simplifies reduced density matrix functional theory (RDMFT). For each symmetry sector,
the sets  $\mathcal{P}^1_N$ and $\mathcal{E}^1_N$ of pure and ensemble $N$-representable one-matrices coincide,
the interaction functionals $\mathcal{F}_{p/e}$ depend only on the natural occupation numbers $\bbf{n}$ and RDMFT therefore reduces \emph{de facto} to a natural occupation number ``functional'' theory.

Those insights have tremendous consequences. Based on Levy's construction \cite{LE79} they allowed us, to discover the \textit{form} of the \emph{exact} functional $\mathcal{F}_p[{\bd{n}}]$ (cf.~\eqref{F}) which differs considerably from the approximate functionals proposed so far \cite{PG16,SKB17}. Intriguingly, $\mathcal{F}_p[{\bd{n}}]$ is given by a bilinear form of square roots (generalizing the two-electron result \cite{LS56}), whose radicants contain \textit{two} terms. The first one is linear in the one-body $N$-representability constraints $\{D^{(j)}({\bd{n}})\}$, while the second summand depends nonlinearly on $\{D^{(j)}({\bd{n}})\}$ and on  the interaction $\hat{V}$ (cf.~Eq.~(\ref{F})). This summand deserves particular attention: First, it arises in the constrained-search \eqref{LevyQ} from those degrees of freedom of $\Psi$ which are \emph{not} determined by the one-matrix. Therefore, it represents within RDMFT \textit{irreducible correlations}, a crucial concept recently established in quantum information theory \cite{LPW02,LW02}. Second, its dependence on $\hat{V}$ emphasizes that the construction of highly accurate functionals based, e.g.,  on  tensor properties \cite{CA00,CGA02} or  $N$-representability conditions for the 2RDM \cite{P05,P17} would necessitate information on the interaction $\hat{V}$, as well.
Third a finite series expansion of that term, $\overline{a}_r\big(\{D^{(i)}({\bbf{n}})\}, \hat{V}\big)$, with respect to $\{D^{(i)}({\bbf{n}})\}$ in conjunction with a fitting scheme would allow one to establish a hierarchy of approximate functionals similar to Jacob's ladder in DFT \cite{PS01}.

Another potentially transformative key result of our work is the discovery of an ``exchange force'' emerging from the fermionic exchange symmetry: The gradient of the exact functional diverges, $\left|\mathrm{d} \mathcal{F}_p/ \mathrm{d} \bbf{n}\right| \sim c_i/\sqrt{D^{(i)}(\bbf{n})}$, as $\bbf{n}$ approaches a facet of the polytope $\mathcal{P}^1_N =\mathcal{E}^1_N$, defined by $D^{(i)}\equiv0$. This repulsive divergence on the boundary of $\mathcal{E}^1_N$ also explains why fermionic occupation numbers $n_k$ \emph{typically} cannot take the extremal values $0$ or $1$. In turn, studying the equation $c_i(\hat{V})=0$ would allow one to systematically identify all (highly non-generic) systems (such as \cite{C18}) for which occupation numbers can be pinned to $0$ or $1$.
It will be one of the crucial future challenges to generalize those new concepts to systems without translational symmetry, with particular focus on \textit{ensemble} RDMFT (i.e., $\mathcal{F}_e$ on $\mathcal{E}^1_N$).

Finally, we would like to stress that all our findings hold for systems with non-fixed particle number, as well and $\hat{V}$ can be \emph{any} (spin-dependent) $p$-particle interaction obeying translational symmetry.

\begin{acknowledgments}
We are grateful to M.\hspace{0.5mm}Piris and coworkers for sharing their data concerning the Hubbard square. We also thank \mbox{P.G.J.\hspace{0.7mm}van Dongen}, \mbox{K.J.H.\hspace{0.5mm}Giesbertz}, I.\hspace{0.5mm}Mitxelena, T.S.\hspace{0.5mm}M\"uller, M.\hspace{0.5mm}Piris and \mbox{R.\hspace{0.5mm}Schade} for helpful comments on the manuscript.
C.S.~acknowledges financial support from the UK Engineering and Physical Sciences Research Council (Grant EP/P007155/1).
\end{acknowledgments}

\bibliography{Refs}

\onecolumngrid
\newpage
\begin{center}\Large{\textbf{Supplemental Material}}
\end{center}
\setcounter{equation}{0}
\setcounter{figure}{0}
\setcounter{table}{0}
\makeatletter
\renewcommand{\theequation}{S\arabic{equation}}
\renewcommand{\thefigure}{S\arabic{figure}}
\vspace{0.5cm}

\twocolumngrid
\section{Derivation of the general form of $\mathcal{F}_p[\emph{n}]$}
\label{sec:GPC}

In a first step, we recall the one-body $N$-representability conditions. We remind the reader that we enumerated the $N$-particle configurations $\bd{q}$ by $r=1,.\ldots,R$.  Then, an $N$-particle state has the form $\ket{\Psi}=\sum^{R}_{r'=1} \alpha_{r'} \ket{r'}$, where $\ket{r'}$ represents a Slater determinant formed from one-particle states $\ket{q}$, $q=1,\ldots,d$.  $d$ is the dimension of the one-particle Hilbert space. Let us consider a single Slater determinant $\ket{r}$, i.e., $\alpha_{r'}=1$ for $r'=r$ and $0$ otherwise.  The corresponding  natural occupation numbers (NONs) are denoted by the vector ${\bbf{v}^{(r)}}$. Its  $q$-th component is $v^{(r)}_q=1$
if $\ket{r}$ contains the one-particle state $\ket{q}$, and otherwise zero.  ${\bbf{v}^{(r)}}$, $r=1,\cdots,R$ are the extremal points of the set $\mathcal{P}^1_N$ of pure-state $N$-representable $\bbf{n}=(n_q)$  and build the vertices of a polytope
$P\equiv \mathcal{P}^1_N$ in the $d$-dimensional space of the NONs. Having determined the vertices one has to find the polytope's facets.
This in general is a nontrivial task.
Each facet, $F_j$, $j=1,\cdots,J$, of $P$  is part of a $(d-1)$-dimensional hyperplane defined by $D^{(j)}({\bbf{n}})=0$ where $D^{(j)}({\bbf{n}})=\kappa^{(j)}_0 + \sum\limits_{k=1}^{d} \kappa^{(j)}_k n_k$. The coefficients, $\kappa^{(j)}_i$, $i=0,1,\cdots,d$, are integers. The polytope is the intersection of the hyperplanes defined by $D^{(j)}({\bbf{n}}) \geq0$,  for all $j$. Therefore, necessary \textit{and} sufficient conditions for the pure-state $N$-representability are the constraints $D^{(j)}({\bbf{n}}) \geq0$ , $j=1,\cdots,J$.

The functions $D^{(j)}({\bbf{n}})$ also allow us to decompose the vertices into two sets.
For given $j$ we decompose the index set $\{1,\cdots ,R\}$ into a set $I_j=\{r_1,\cdots,r_j\}$ and its complement  such that  ${\bbf{v}}^{(r)} \in F_j$ for $r \in I_j$ and ${\bbf{v}}^{(r)} \notin F_j$
otherwise. It is $D^{(j)}({\bbf{v}}^{(r)})=0$ for $r \in I_j$  and $D^{(j)}({\bbf{v}}^{(r)}) > 0$ for $r \notin I_j$ \cite{K09,SBV17}.

What remains is  the derivation of the relation between $\{D^{(j)}({\bbf{n}})\}$ and $\{|\alpha_r|^2\}$.
Each function $D^{(j)}({\bbf{n}})$ determines an operator $D^{(j)}(\hat{{\bbf{n}}})$ with $\hat{n}_q=c^\dag_q c_q $. Since $\langle \Psi|\hat{n}_q|\Psi\rangle=n_q$ we have $\langle \Psi|D^{(j)}(\hat{{\bbf{n}}}) |\Psi\rangle=D^{(j)}({\bbf{n}})$. On the other hand we can substitute $|\Psi\rangle=\sum^{R}_{r'=1} \alpha_{r'}|r'\rangle$ on its l.h.s.. With $({\bf{A}})_{jr} \equiv A_{jr} := D^{(j)}({\bbf{v}}^{(r)})$
 this leads to

\begin{equation} \label{eqA1}
D^{(j)}({\bbf{n}}) = \sum^{R}_{r'=1} A_{jr'} |\alpha_{r'}|^2 \\ \ .
\end{equation}
This equation establishes a relation between the NONs and $\{ |\alpha_r|^2\}$ involving the functions $\{D^{(j)}({\bbf{n}})\}$ which define the domain of pure-state representability.

For the constrained minimization of $\langle \Psi|\hat{V}|\Psi\rangle= \sum_{r,r'} V_{rr'}
\alpha^{*}_r \alpha_{r'}$ for fixed ${\bbf{n}}$ we assume for a moment that
 $V_{rr'}=- \tilde{\eta}_r \tilde{\eta}_{r'}|V_{rr'}|$ for $r \neq r'$, as well as,  real coefficients $\alpha_r= \eta_r|\alpha_r|$) with $\tilde{\eta}_r =\pm 1$ and  $\eta_r =\pm 1$. In that case the  minimization with respect to the phase factors  $\{\eta_r \}$ is accomplished by the choice $\tilde{\eta}_r \equiv \eta_r $. Then the expectation value
  $\langle \Psi|\hat{V}|\Psi\rangle$ takes the form

 \begin{equation} \label{eqA1a}
\tilde{\mathcal{F}} [\{|\alpha_r|\}] = \sum_{r} V_{rr}  |\alpha_r|^2 - \sum_{r \neq r'}  |V_{rr'}|  |\alpha_r| |\alpha_{r'}|\\.
\end{equation}
 To derive $\mathcal{F}[\bd{n}]$ we have to determine $\{|\alpha_r|\}$ as a function of $\bbf{n}$.
 This can be done as follows. Introducing the symmetric and semi-definite  matrix ${\bf{C}}={\bf{A}}^t \bf{A}$ and  operating with ${\bf{A}}^t$ on Eq.~(\ref{eqA1}) one obtains $\sum^{J}_{j=1} ({\bf{A}}^t)_{rj}D^{(j)}({\bbf{n}})= \sum^{R}_{r'=1} C_{rr'} |\alpha_{r'}|^2$. If $d=R$, there are as many NONs as  coefficients $\{|\alpha_r|\}$. In that case $\{ |\alpha_r|\}$ is uniquely determined by $\{D^{(j)}({\bbf{n}})\}$, i.e. by the NONs, ${\bbf{n}}$. This always holds if the poytope $P$ is a simplex, which was the case for the example of three fully polarized electrons on a ring of six lattice sites. Note that there are $J$ functions
 $\{D^{(j)}({\bbf{n}})\}$ and only
$d \leq J$ NONs.  In  case $d < R$ (occurs for  $L$ and $N$ large enough),  however,  $\{|\alpha_r|\}$ are not uniquely determined by the NONs.
In that case $\bf{C}$ has zero-eigenvalues, i.e., its rank, $d$, is smaller  than  $R$. Let $\{{\bf{w}}^{(l)}\}$ be the eigenvectors of $\bf{C}$ and $\{c_l\}$ its corresponding eigenvalues. $c_l > 0$ for $l=1,\cdots,d$ and
$c_l = 0$ for $l=d+1,\cdots,R$. Substituting the expansion

\begin{equation}  \label{eqA2b}
( |\alpha_1|^2,\cdots, |\alpha_{R}|^2)= \sum^{R}_{l=1} a^{(l)} {\bf{w}}^{(l)}
\end{equation}
into the equation $\sum^{J}_{j=1} ({\bf{A}}^t)_{rj}D^{(j)}({\bbf{n}})= \sum^{R}_{r'=1} C_{rr'} |\alpha_{r'}|^2$ from above and taking the orthonormality of $\{{\bf{w}}^{(l)}\}$ into account allows us to determine $a^{(l)}$ for $l=1,\ldots,d$. This yields $a^{(l)}=\sum^{d}_{l=1} \big(c^{-1}_l {\bf{w}}^{(l)t} {\bf{A}}^t {\bf{D}}({\bbf{n}}) \big) w_r^{(l)}$
for $l=1,\ldots,d$. Substituting these $a^{(l)}$ into the r.h.s. of  Eq.~(\ref{eqA2b}) we arrive at

\begin{eqnarray}  \label{eqA2}
|\alpha_{r}|({\bf{n}},{\bf{a}})
&=&\Big[\sum^{J}_{j=1}b^{(j)}_rD^{(j)}({\bf{n}}) + \sum^{R}_{l=d+1} a^{(l)} w_r^{(l))} \Big]^{1/2}\!\!\!.
\end{eqnarray}
The coefficients $\{b^{(j)}_r\}$ follow from
$\sum^{d}_{l=1} \big(c^{-1}_l {\bf{w}}^{(l)t} {\bf{A}}^t {\bf{D}}({\bbf{n}}) \big) w_r^{(l))}=\sum^{J}_{j=1} b^{(j)}_rD^{(j)}({\bbf{n}})$ where  ${\bf{D}}({\bbf{n}})=(D^{(1)}({\bbf{n}}),\cdots,D^{(J)}({\bbf{n}}))^t$.
The absolute values $\{|\alpha_r|\}$ are fixed by the NONs through $\{D^{(j)}({\bbf{n}})\}$ and by the independent real variables ${\bf{a}}=(a^{(d+1)},\ldots, a^{(R)})$.

To get $\mathcal{F}_p$ the result Eq.~(\ref{eqA2}) has to be substituted into  Eq.~(\ref{eqA1a}) with a subsequent minimization with respect to ${\bf{a}}$. This is
a nontrivial problem which in general can not be performed analytically. Substituting its solution ${\bf{a}}(\{D^{(j)}({\bbf{n}})\},\hat{V})$ into Eq.~(\ref{eqA2}) and this expression into Eq.~(\ref{eqA1a}) yields the functional

\begin{widetext}
\begin{equation} \label{eqA3}
\mathcal{F}_p[{\bbf{n}}]= \sum^{R}_{r,r'=1}V_{rr'} \sqrt{\sum^{J}_{j=1}\Big[b^{(j)}_rD^{(j)}({\bbf{n}})+\overline{a}_r(\{D^{(j)}({\bbf{n}})\}, \hat{V})\Big]} \,\,
\sqrt{\sum^{J}_{j=1}\Big[b^{(j)}_{r'}D^{(j)}({\bbf{n}})+\overline{a}_{r'}(\{D^{(j)}({\bbf{n}})\},\hat{V})\Big]}  \
\end{equation}
\end{widetext}
where $V_{rr'}=-|V_{rr'}|$ for all $r \neq r'$ and $\overline{a}_r(\{D^{(j)}(\bbf{n})\},\hat{V})
=\sum^{R}_{l=d+1} \, a^{(l)}(\{D^{(j)}(\bbf{n})\},\hat{V}) \, w^{(l)}_r$. Note, the dependence on $\hat{V}$
occurs through the matrix elements $\{V_{rr'}\}$ of $\hat{V}$.
\\

The result (\ref{eqA3}) simplifies for ${\bbf{n}}$ close to a facet $F_j$. Remember that $D^{(j)}({\bbf{n}}) \to 0$ for ${\bbf{n}} \to F_j$. As described above we can decompose the set $r=1,\ldots,R$ of the  vertex-indices  into two subsets, $I_j$ and its complement. Then it follows from Eq.~(\ref{eqA1})
that $ |\alpha_r |= \sqrt{D^{(j)}({\bbf{n}})} \, {\beta}_{r}$ for all $r \notin I_j$ and
 $|\alpha_r|=|\alpha^{(j)}_r| + \mathcal{O}(D^{(j)}({\bbf{n}})) $ for all $r \in I_j$.  The real and non-negative variables $\bd{\beta}=(\beta_{r})$ have to fulfil
 $\sum_{r' \notin I_j} A_{jr'}({\beta}_{r'})^2 =1$, which follows from Eq.~(\ref{eqA1}).
  $\{\alpha_r^{(j)}\}$ are the coefficients of the normalized $N$ particle state $|\Psi^{(j)}\rangle= \sum_{r' \in I_j} \alpha_{r'}^{(j)} |r'\rangle$
 build from Slater determinants $|r\rangle$ corresponding  to the vertices of the facet $F_j$, only. Substituting these quantities into Eq.~(\ref{eqA1a}) yields

\begin{eqnarray} \label{eqA4}
\tilde{\mathcal{F}} [\{|\alpha_r|\}] &=& \tilde{\mathcal{F}} [\{|\alpha_r^{(j)|}\}] + \nonumber\\
&+& \delta\tilde{\mathcal{F}} [\{|\alpha^{(j)}_r|\},\bd{\beta}]\sqrt{D^{(j)}({\bbf{n}})} + \nonumber\\
&+& \mathcal{O}(D^{(j)}({\bbf{n}}))  \ ,
\end{eqnarray}
with
\begin{equation} \label{eqA5}
\tilde{\mathcal{F}} [\{|\alpha^{(j)}_r|\}] = \sum_{r,r' \in I_j} V_{rr}  |\alpha^{(j)}_r|  |\alpha^{(j)}_{r'}|
\end{equation}
and
\begin{equation} \label{eqA6}
\delta\tilde{\mathcal{F}} [\{|\alpha^{(j)}_r|\},\bd{\beta}] = 2\sum_{r \in I_j ,r' \notin I_j} V_{rr'}  |\alpha^{(j)}_r|  \, \beta_{r'}\\.
\end{equation}
Again, it is $V_{rr'}=-|V_{rr'}|$  for all $r \neq r'$.

$\tilde{\mathcal{F}} [\{|\alpha^{(j)}_r|\}]$ is like   $\tilde{\mathcal{F}} [\{|\alpha_r|\}]$ but restricted to a subspace spanned by all basis  states $|r\rangle$ with $r \in I_j$.
Its minimization with respect to $\{|\alpha^{(j)}_r|\}$ has to  be performed in analogy to that of  $\mathcal{F}_p [\{|\alpha_r|\}]$, but now  under the constraint   ${\bbf{n}^{(j)}}$ fixed.  $\bbf{n}^{(j)}$  is a chosen reference point in
$F_j$ which is the limiting point of $\bd{n} \to F_j$. This minimization process yields  $\{|\alpha^{(j)}_r|({\bbf{n}^{(j)}})\}$ and finally
 $\mathcal{F}_p [\{|\alpha^{(j)}_r|({\bbf{n}^{(j)}})\}] = \mathcal{F}^{(j)}_p [{\bbf{n}^{(j)}}]$. Note, the dependence of
 $\{|\alpha^{(j)}_r|\}$ on $\{V_{r,r'}\}$, $r,r' \in I_j$ is suppressed.
 Furthermore, $ \{|\alpha^{(j)}_r|\}$
 in the second line of Eq.~(\ref{eqA4}) has to be replaced by $\{|\alpha^{(j)}_r|({\bbf{n}^{(j)}})\}$.
 It remains the minimization of  $\delta\tilde{\mathcal{F}}$ with respect to $\bd{\beta}$ which yields
$\bd{\beta}({\bbf{n}^{(j)}})$ where the dependence on $\{V_{rr'}\}$ is suppressed, as well.
This completes the minimization of  $\tilde{\mathcal{F}} [\{|\alpha_r|\}]$
for $\bf{n}$ approaching  ${\bf{n}}^{(j)}$ in  $F_j$. The  functional takes the final form

\begin{widetext}
\begin{equation} \label{eqA7}
\mathcal{F}_p [{\bbf{n}}] = \mathcal{F}^{(j)}_p [{\bbf{n}^{(j)}}]
-2\sum_{r \in I_j ,r' \notin I_j} |V_{rr'}| \,|\alpha^{(j)}_r|({\bbf{n}^{(j)}}) \, \beta_{r'}({\bbf{n}^{(j)}}) \sqrt{D^{(j)}({\bbf{n}})}
+ \mathcal{O}(D^{(j)}({\bbf{n}}))
\end{equation}
\end{widetext}

In case that the interaction matrix does \emph{not} have the form $V_{rr'} =- \tilde{\eta}_r \tilde{\eta}_{r'}|V_{rr'}|$ we have to minimize the functional
\begin{equation}\label{eqA2a}
\tilde{\mathcal{F}}[\{|\alpha_r|\},\bd{\eta}] = \sum_{r,r'}  V_{rr'} \eta_r^\ast \,\eta_{r'} |\alpha_r| |\alpha_{r'}|  \ ,
\end{equation}
where $\bd{\eta}=\{\eta_r\}$ again are the phase factors of $\{\alpha_r\}$.
Similar as above , for fixed  $(\bd{n},\bd{\eta})$  we require additional parameters   ${\bf{a}}=(a^{(l)})$ in order to fix $\{|\alpha_r|\}$. Performing the minimization on the r.h.s. of Eq.~(\ref{eqA2a}) with respect to $\bf{a}$
yields $a^{(l)}(\{D^{(j)}({\bbf{n}})\},\hat{V}, \bd{\eta})$ and $\{|\alpha_r|\}(\{D^{(j)}({\bbf{n}})\},\hat{V}, \bd{\eta})$
follows from Eq.~(\ref{eqA2}) by substituting  $a^{(l)}(\{D^{(j)}({\bbf{n}})\},\hat{V}, \bd{\eta})$. Then, substitution of
$\{|\alpha_r|\}(\{D^{(j)}({\bbf{n}})\},\hat{V}, \bd{\eta})$ into Eq.~(\ref{eqA2a}) yields
$\tilde{\mathcal{F}}[\bd{n},\bd{\eta}]$. The final step concerns the minimization with respect to the Ising-like variables
$\bd{\eta}$. Let $ \overline{\bd{\eta}}(\{D^{(j)}({\bbf{n}})\},\hat{V})$ denote the minimizing phase factors. The substitution of those
into $\tilde{\mathcal{F}}[\bd{n},\bd{\eta}]$ yields the final result for  $\mathcal{F}_p$ given in Eq.~(9) of the main text with
$\overline{a}_r\big(\{D^{(j)}({\bbf{n}})\},\hat{V}\big)=\sum^{R}_{l=d+1}\, a^{(l)}\big(\{D^{(j)}({\bbf{n}})\},\hat{V}, \overline{\bd{\eta}}(\{D^{(j)}({\bbf{n}})\},\hat{V})\big) w^{(l)}_r$.

\subsection{Derivation of the exchange force for the case of $\mathcal{P}_N^1$ being a simplex}
In the case where $\mathcal{P}_N^1$ takes the form of a simplex, the derivation of the exchange force is apparently much easier (cf.~Eq.~(6)) than for the general case of an arbitrary polytope $\mathcal{P}_N^1= \mathcal{E}_N^1$. We prove in the following that this exchange force is repulsive in the sense that it repels $\bbf{n}$ from the boundary of $\mathcal{P}_N^1$. For this, we revisit Levy's construction where we use again the ansatz $\ket{\Psi}=\sum_{r=1}^R \eta_r |\alpha_r|$ and assume that the interaction matrix elements $V_{r r'}$ and therefore also the phases factors $\eta_r$ are real-valued, i.e., $\eta_r =\pm 1$. As in the main text, we label the one-body $N$-representability constraints $D^{(r)}(\bbf{n})\geq 0$ such that the respective facet does not contain the vertex $\bd{v}_r$, i.e.~we have $D^{(r)}(\bd{v}_{r'})=0$ whenever $r\neq r'$. For simplicity, we ``normalize'' each $D^{(r)}\geq 0$ such that $D^{(r)}(\bd{v}_r) =1$. Moreover, we recall Eq.~(5), i.e.
\begin{equation}\label{DvsAsimplex}
D^{(r)}(\bbf{n}) = |\alpha_r|^2\,.
\end{equation}
Let us now consider $\bbf{n}$ very close, in a distance $\varepsilon$ to the facet described by $D^{(s)}\equiv0$ and assume that the distances $D^{(r)}(\bbf{n})$, to all other facets are much larger, i.e., $D^{(r)}(\bbf{n})\gg D^{(s)}(\bbf{n})\equiv \varepsilon$ for all $r\neq s$. W.l.o.g.~we assume $s=1$. Resorting to Levy's construction and the general ansatz for $\ket{\Psi}$, we find
\begin{widetext}
\begin{eqnarray}
\mathcal{F}_p[\bbf{n}] &=& \min_{\{\eta_r\}} \sum_{r,r'=1}^R \eta_r \eta_{r'} V_{r r'} \sqrt{D^{(r)}(\bbf{n})D^{(r')}(\bbf{n})} \nonumber \\
&=& \min_{\{\eta_r\}_{r>1}} \min_{\eta_1}\left[\sum_{r,r'>1} \eta_r \eta_{r'} V_{r r'} \sqrt{D^{(r)}(\bbf{n})D^{(r')}(\bbf{n})} +
2 \sum_{r>1} \eta_r \eta_1 V_{r 1} \sqrt{D^{(r)}(\bbf{n})D^{(1)}(\bbf{n})} +
V_{1 1} D^{(1)}(\bbf{n})
\right]\nonumber \\
&=& \min_{\{\eta_r\}_{r>1}}\left[\sum_{r,r'>1} \eta_r \eta_{r'} V_{r r'} \sqrt{D^{(r)}(\bbf{n})D^{(r')}(\bbf{n})} -
2 \sqrt{D^{(1)}(\bbf{n})}\,\left|\sum_{r>1} \eta_r  V_{r 1} \sqrt{D^{(r)}(\bbf{n})} \right| +
V_{1 1} D^{(1)}(\bbf{n})
\right]\,.
\end{eqnarray}
\end{widetext}
We remind the reader that the one-body $N$-representability constraints read $D^{(r)}({\bbf{n}})=\kappa^{(r)}_0 + \sum^{d}_{q=1}\kappa^{(r)}_qn_q \geq 0$.  The gradient $\nabla_{{\bd{n}}} \mathcal{F}_p[\bbf{n}]$  contains  products of
  $\partial \mathcal{F}_p / \partial D^{(r)}$ and  $\nabla_{{\bd{n}}}D^{(r)}({\bbf{n}})$. The latter equals the vector
  ${\bd{\kappa}^{(r)}}=(\kappa^{(r)}_1,\ldots,\kappa^{(r)}_d)^t$, which is anti-parallel to the normal vector of the corresponding facet.
 Taking now the gradient of $\mathcal{F}_p[\bbf{n}]$, only the term in the middle yields a contribution which diverges in the limit $\varepsilon \rightarrow 0^+$ (since we assumed $D^{(r)}(\bbf{n})\gg D^{(1)}(\bbf{n})\equiv \varepsilon$ for all $r>1$).
 $\partial \mathcal{F}_p / \partial D^{(1)}$  is proportional to $1/\sqrt{D^{(1)}(\bbf{n})}$ which is positive,
 and its prefactor $-\,\left|\sum_{r>1} \eta_r  V_{r 1} \sqrt{D^{(r)}(\bbf{n})} \right|$ is apparently negative.
 Consequently, the  exchange force ${\bd{f}}_{ex}(\bd{n}) = - \nabla_{{\bd{n}}} \mathcal{F}_p[\bbf{n}]$ is parallel
 to ${\bd{\kappa}^{(1)}}$, i.e., it points towards the interior of the polytope. Hence, the exchange force is repulsive in the sense that it repels  $\bbf{n}$ from the polytope's boundary.

\section{Proof of $ \mathcal{F}_p =\mathcal{F}_e$}
\label{sec:Proof}
We assume $\alpha_r= \eta_r |\alpha_r |$ in  $|\Psi\rangle = \sum^{R}_{r=1}\alpha_r |r\rangle $ to be real and that the  interaction matrix elements are of the form $V_{rr'} \equiv \langle r |\hat{V}|r'\rangle= - \tilde{\eta}_r \tilde{\eta}_{r'}| V_{rr'}|$  for all $r \neq r'$.  $\eta_r= \pm 1$ and $\tilde{\eta}_r= \pm 1$.  Then the minimization of the expectation value $\langle \Psi |\hat{V}|\Psi\rangle$ with respect to the phase factors $\{\eta_r\}$ is done for $\eta_r \equiv \tilde{\eta}_r$ leading to
\begin{eqnarray} \label{eqB1}
\tilde{\mathcal{F}}[\{|\alpha_r|\}]&=& \min_{\{\eta_r\}} \langle \Psi |\hat{V}|\Psi\rangle  \nonumber\\
&=&\sum^{R}_{r=1} V_{rr}  |\alpha_r|^2 - \sum^{R}_{r \neq r'=1}  |V_{rr'}|  |\alpha_r| |\alpha_{r'}|  \,  .
\end{eqnarray}

Choose an $N$-particle ensemble $ \hat{\Gamma} = \sum^{R}_{r,r'=1} \Gamma_{rr'} |r \rangle \langle r'|$ . Then it follows $\langle \hat{V} \rangle_{\hat{\Gamma}}=Tr_N(\hat{V}\hat{\Gamma})=\sum^{R}_{r=1} V_{rr} \Gamma_{rr} - \sum^{R}_{r \neq r'=1} \tilde{\eta}_r \tilde{\eta}_{r'} |V_{rr'}| \Gamma_{rr'}$. A necessary condition for $\hat{\Gamma} \geq 0$
is $|\Gamma_{rr'}|^2 \leq \Gamma_{rr}\Gamma_{r'r'}$ for all $r \neq r'$. The choice $\Gamma_{rr'} =\tilde{\eta}_r \tilde{\eta}_{r'}\sqrt{\Gamma_{rr}\Gamma_{r'r'}}$ minimizes  $\langle \hat{V} \rangle_{\hat{\Gamma}}$ for \textit{fixed} diagonal elements $\{\Gamma_{rr}\}$ and leads to $\min_{\{\Gamma_{r r'}\}_{r \neq r'}} \langle \hat{V} \rangle_{\hat{\Gamma}} \equiv \tilde{\mathcal{F}} [\{\sqrt{\Gamma_{rr}}\}]$.
This choice also implies  $\hat{\Gamma}= |\Phi \rangle  \langle \Phi|$ with $ |\Phi \rangle = \sum^{R}_{r=1} \tilde{\eta}_r \sqrt{\Gamma_{rr}} \, |r \rangle$, i.e. the corresponding $N$-particle density operator, $\hat{\Gamma}$, is positive semi-definite. Final minimization of $\tilde{\mathcal{F}}\{|\alpha|_r\}$ and  $\tilde{\mathcal{F}}\{\sqrt{\Gamma_{rr}}\}$ with respect to $\{|\alpha|_r\}$ and
$\{\sqrt{\Gamma_{rr}}\}$ under  constraints $\bd{n}=\{n_q\}$ fixed, leads to $\mathcal{F}_p=\mathcal{F}_e \equiv \mathcal{F}$.

\section{Derivation of $\mathcal{F}_p[\emph{n}]$ for $N$=3 fully polarized electrons in one dimension and $L$=6}
\label{sec:Full Polarized}

The one-particle momenta $k=(2\pi/6) \, \nu$ from the first Brillouin zone are chosen as $\nu=0,1,2,3,4,5$.  Taking only nearest neighbor hopping into account  this leads to the one-particle energies $\varepsilon_\nu=-2t \cos (2 \pi \nu/6)$.
$t >0$ is the nearest neighbor hopping parameter. The ground state for noninteracting spinless electrons ( i.e., $\hat{V} \equiv 0$) is $\ket{\nu_1,\nu_2,\nu_3}^{(0)}=|0,1,5 \rangle$ for which the total momentum is $K=(2\pi/6))(1+5)$(mod$6$)$=0$. If $|\langle {\bd{q}} |\hat{V}| {\bd{q}'}  \rangle| $ for
the $3$-particle states $| {\bd{q}} \rangle=  c^{\dagger}_{q_1\uparrow}c^{\dagger}_{q_2\uparrow}c^{\dagger}_{q_3\uparrow} |0\rangle$ is below a critical value for all $ {\bd{q}} $, $ {\bd{q}'} $ with $K=0$ the ground state of the interacting system will stay in this symmetry sector. It is easy to show that the zero-momentum space is spanned by four states $\ket{\nu_1,\nu_2,\nu_3}=|0,1,5 \rangle,$ $|0,2,4 \rangle$, $|1,2,3 \rangle$ and $|3,4,5 \rangle$, denoted by $|r\rangle$, $r=1, \cdots, 4$. Then a general three-particle state in this symmetry sector is represented as
$|\Psi\rangle=\sum\limits_{r=1}^4 \alpha_r |r\rangle$.
 Note that the number, $d\equiv L=6$, of one-particles states $|q \rangle$  is larger than,  $R=4$, the dimension of the three-particle subspace. This implies besides the normalization $\sum^{5}_{\nu=0} n_{\nu} =3$ additional identities for the NONs ($n_{\nu}$)  independent on $\{\alpha_r\}$. Since fully polarized electrons correspond to spinless fermions, the spin variables are suppressed.

 It is straightforward to determine
 $\{n_{\nu}\}$ as a function of $\{\alpha_r\}$. From this relation and the normalization condition for $\{\alpha_r\}$ one obtains

  \begin{equation} \label{eqD1}
n_{3}=1-n_0, \,\, n_ {4}=1-n_1, \,\, n_5=1-n_2 \quad.
\end{equation}

Accordingly, there are three-independent NONs, only. We choose ${\bbf{n}}=(n_0, n_1,n_2)$.

Now one could follow the general scheme described in the main text to determine the vertices of the polytope and then the facets which yields the functions $\{D^{(j)}({\bbf{n}})\}$. Since for the present case the set of linear equations
relating  $\{n_{\nu}\}$ and  $\{\alpha_r\}$ is rather simple one can solve this set directly. One obtains for $r=1,\cdots,4$
\begin{equation} \label{eqD2}
 |\alpha_r| = \sqrt{D^{(r)}({\bbf{n}})/2} \,.
\end{equation}
with
\begin{eqnarray} \label{eqD3}
&& D^{(1)}({\bbf{n}})= n_0+n_1-n_2  \nonumber\\
&& D^{(2)}({\bbf{n}})= n_0-n_1+n_2 \nonumber\\
&&  D^{(3)}({\bbf{n}})=2 - n_0-n_1- n_2 \nonumber\\
&& D^{(4)}({\bbf{n}})=-n_0+n_1+n_2
\end{eqnarray}
The validity of $\sum\limits_{r=1}^4 |\alpha_r|^2=1$ is obvious. Eq.~(\ref{eqD3}) ist identical to Eq.~(3) of the main text.

$ |\alpha_r| \geq 0$ and Eq.~(\ref{eqD2}) yields  the generalized  constraints, $D^{(j)} ({\bbf{n}}) \geq 0$,\, $j=1,\ldots,4$  on ${\bbf{n}}$. These guarantee that ${\bbf{n}}$ is pure-state $N$-representable, in the
sector $K=0$. They define four planes building a three-dimensional polytope (a tetrahedra, which is a simplex). This polytope is identical to that of the so-called Borland-Dennis setting for three spinless fermions in a six-dimensional one-fermion Hilbert space without any symmetry conditions \cite{BD72,R07}.
Substituting $\{ |\alpha_r|\}$ from Eq.~(\ref{eqD2}) into
$\bra{\Psi}\hat{V}\ket{\Psi}=\sum_{r,r'} V_{rr'}\eta_r \eta_{r'} \,  |\alpha_r| \, |\alpha_{r'}|$ and minimizing with respect to $\{\eta_r\}$ one obtains the final result which is of the form of Eq.~(6) with $\{D^{(j)}(\bd{n})\}$ from Eq.~(\ref{eqD3}).

\section{Derivation of $ \mathcal{F}[\emph{n}]$ for the Hubbard-square, $N=4$, $L=4$, $K=2(2\pi/4)$, $S=0$  and parity $p=-1$}
\label{sec:Derivation}

To determine all Slater determinants $|k_1 m_1, k_2 m_2, k_3 m_3, k_4 m_4 \rangle$ with total momentum
$K=\sum\limits_{n=1} ^4 k_n ({\rm mod}\, 2\pi) =  2\pi/4 \,\sum\limits_{n=1} ^4 \nu_n ({\rm mod}\, 4) =2 \cdot  2\pi/4$ and total magnetization $M_z=\sum\limits_{n=1}^4 m_n=0$ ($m_n = \pm 1/2$) is straightforward. With $\nu_n \in \{0,1,2,3\}$ skipping $2\pi/4$ and use of $+1/2=\uparrow ,\, -1/2=\downarrow$ one obtains ten states

\begin{eqnarray} \label{eqC1}
&& |0\uparrow, 0\downarrow, 3 \uparrow, 3\downarrow \rangle, \ |0\uparrow, 0\downarrow, 1 \uparrow, 1\downarrow \rangle, \nonumber\\
&&  |2 \uparrow, 2\downarrow,3\uparrow, 3\downarrow \rangle,  |1\uparrow, 1\downarrow, 2 \uparrow, 2\downarrow \rangle,\nonumber\\
&&  |0\downarrow, 1 \uparrow, 2\downarrow, 3\uparrow  \rangle,  | 0\uparrow, 1 \downarrow, 2\uparrow, 3\downarrow \rangle,\nonumber\\
&&  0\downarrow, 1 \downarrow, 2\uparrow, |3\uparrow \rangle,  |0\uparrow, 1 \uparrow, 2\downarrow, 3\downarrow \rangle, \nonumber\\
&&   |0\uparrow, 1 \downarrow, 2\downarrow, 3\uparrow \rangle,  0\downarrow, 1 \uparrow, 2\uparrow, |3\downarrow \rangle \ .
\end{eqnarray}

Due to the isotropy in spin space and the reflection symmetry $P : i \to L-i+1$ implying $P : \nu \to - \nu (mod L)$ all basis states can be chosen to be eigenstates of the operator of the total spin squared, $\vec{\hat{S}}^2$, and the parity operator $\hat{P}$ with eigenvalues $S(S+1)$ and $p= \pm 1$, respectively. The ground state for zero interactions is two-fold degenerate. The degeneracy is lifted in first order in $U$. The corresponding groundstate for $U=0^{+}$ is given by $\frac{1}{\sqrt{2}}[ |0\uparrow, 0\downarrow, 3 \uparrow, 3\downarrow \rangle -  |0\uparrow, 0\downarrow, 1 \uparrow, 1\downarrow \rangle]$ which is an eigenstate of $\vec{S}^2$ and  $\hat{P}$ with eigenvalues $0$ and $p= - 1$, respectively.

Then we get for  $S=0$ and $p=-1$ the following three basis states

\begin{eqnarray} \label{eqC2}
&& \frac{1}{\sqrt{2}}\Big[  0\uparrow, 0\downarrow,|3 \uparrow, 3\downarrow \rangle - |0\uparrow, 0\downarrow, 1 \uparrow, 1\downarrow \rangle \Big] \ ,  \nonumber\\
&& \frac{1}{\sqrt{2}}\Big[ |1 \uparrow, 1 \downarrow, 2 \uparrow, 2 \downarrow \rangle + | 2\uparrow, 2\downarrow,3\uparrow, 3\downarrow \rangle \Big]  , \nonumber\\
&& \frac{1}{4\sqrt{3}} \Big[-2 \, \big( |0\downarrow, 1 \uparrow, 2\downarrow,3\uparrow \rangle +  | 0\uparrow, 1 \downarrow, 2\uparrow,3\downarrow \rangle \big) \nonumber\\
&& + \big( 0\downarrow, 1 \downarrow, 2\uparrow, |3\uparrow \rangle + |0\uparrow, 1 \uparrow, 2\downarrow, 3\downarrow \rangle \big) \nonumber\\
&& + \big( | 0\uparrow, 1 \downarrow, 2\downarrow, 3\uparrow \rangle +  |0\downarrow, 1 \uparrow, 2\uparrow, 3\downarrow,  \rangle \big)\Big]  \ ,
\end{eqnarray}
which will be denoted by $|r \rangle$ , $r= 1, \cdots, 3$. With $| \Psi\rangle = \sum\limits^3_{r=1} \alpha_r |r\rangle$ it is straightforward  to express the NONs  by $\{|\alpha_r|^2 \}$ :

\begin{eqnarray} \label{eqC3}
&& n_{0 \uparrow} = |\alpha_1|^2 + \frac{1}{2} |\alpha_3|^2 \nonumber\\
&& n_{1 \uparrow} = \frac{1}{2} [|\alpha_1|^2 + |\alpha_2|^2 +|\alpha_3|^2] = \frac{1}{2}\nonumber\\
&& n_{2 \uparrow} =  |\alpha_2|^2 +  \frac{1}{2} |\alpha_3|^2 \nonumber\\
&& n_{3 \uparrow} =\frac{1}{2} [|\alpha_1|^2 + |\alpha_2|^2 +|\alpha_3|^2] = \frac{1}{2} \ ,
\end{eqnarray}
where the normalization $\sum\limits^3_{r=1} |\alpha_r|^2 =1$ was used. Due to $S=0$ it is
$n_{\mu \uparrow} =n_{\mu \downarrow}$.  Furthermore Eq. (\ref{eqC3}) implies
$n_{0 \uparrow}+n_{2 \uparrow}=1$.  With  $n_{\mu} \equiv n_{\mu \uparrow} =n_{\mu \downarrow}$ it follows from Eq.~(\ref{eqC3})

\begin{eqnarray} \label{eqC4}
&& n_{0} =|\alpha_1|^2 + \frac{1}{2} |\alpha_3|^2 = 1 -n_2\nonumber\\
&& n_{2} =|\alpha_2|^2 + \frac{1}{2} |\alpha_3|^2\nonumber\\
&& n_{1} = n_{3} = 1/2 \ .
\end{eqnarray}

Accordingly there is one independent NON, only. We choose  $n_2$ and  identify ${\bbf{n}}$ with $n_2$ , being restricted to $0 \leq n_2 \leq 1$.
Therefore the ``facets'' are defined be $D^{(1)}({\bbf{n}})=1-n_2=0$ and $D^{(2)}({\bbf{n}})=n_2=0$

The matrix $(A_{jr}) \equiv (D^{(j)}({\bbf{v}}^{(r)}))$ in Eq.~(\ref{eqA1})  becomes

\begin{eqnarray} \label{eqC5}
(A_{jr})&=&  \frac{1}{2}\left(
\begin{array}{ccc}
2 & 0 &1\\
0 & 2&1 \\
 \end{array}
\right)  \  ,
\end{eqnarray}
which leads to (see part I)

\begin{eqnarray} \label{eqC6}
(C_{rr'})&=&  \frac{1}{4} \left(
\begin{array}{ccc}
4 & 0& 2 \\
0 & 4& 2 \\
2 & 2& 2
 \end{array}
\right)  \  .
\end{eqnarray}
The eigenvalues $\{c_l\}$and the corresponding orthonormalized eigenvectors $\{(w^{(l)}_r)\}$ are $c_1=1 \ , c_2=3/2 \ ,c_3=0 $ and $(w^{(1)}_r)=(1/\sqrt{2})(1,-1,0)^t  \ , (w^{(2)}_r)=(1/\sqrt{3})(1,1,1)^t  \ , (w^{(3)}_r)=(1/\sqrt{6})(1,1,-2)^t$, respectively. Substituting these expressions into Eq.~(\ref{eqA2b}) yields

\begin{eqnarray} \label{eqC7}
\begin{pmatrix}
|\alpha_1|^2\\
|\alpha_2|^2\\
|\alpha_3|^2\\
\end{pmatrix}
=
\begin{pmatrix}
D^{(1)}({\bbf{n}}) +(a^{(3)}-\frac{1}{\sqrt{6}})w^{(3)}_1\\
D^{(2)}({\bbf{n}}) +(a^{(3)}-\frac{1}{\sqrt{6}})w^{(3)}_2\\
(a^{(3)}-\frac{1}{\sqrt{6}})w^{(3)}_3
\end{pmatrix}  \ .
\end{eqnarray}
\\
It is straightforward to calculate the matrix elements  $V_{rr'}=\langle r|\hat{V} |r'  \rangle $ of the Hubbard interaction $\hat{V}=U \sum^4_{i=1} \hat{n}_{i\uparrow}\hat{n}_{i\downarrow}$. As a result one obtains

\begin{eqnarray} \label{eqC8}
(V_{rr'})&=&(U/4) \left(
\begin{array}{ccc}
3 &-1&  -\sqrt{6}\\
-1 & 3& -\sqrt{6}\\
  -\sqrt{6}& -\sqrt{6}&6\\
\end{array}
\right)  \ .
\end{eqnarray}
Note, For $U >0$ all \textit{nondiagonal} elements are negative. Therefore, the Hubbard interaction belongs to the class of pair interactions for which Eqs.~(\ref{eqA3}) and (\ref{eqA7})  hold. Therefore it is $\mathcal{F}_p[\bbf{n}]=\mathcal{F}_e[\bbf{n}] \equiv \mathcal{F}[\bbf{n}]$

 To get the functional $\mathcal{F}[\bbf{n}]$  we have to minimize $\sum^3_{r,r'=1} V_{rr'}\alpha_r\alpha_{r'}$
with respect to the single independent degree of freedom, $a^{(3)}$, and have to follow the scheme described in part I . This leads to $\mathcal{F}[\bbf{n}]$ of the form of Eq.~(9). Here we illustrate another route where first   $(V_{rr'})$ is diagonalized. The eigenvalues of $(V_{rr'})$ are $U(1,2,0)$ with corresponding orthonormalized eigenvectors $\vec{v}_1=(1/\sqrt{2})(1,-1,0)^t$, $\vec{v}_2=1/(2\sqrt{2})(1,1,-\sqrt{6})^t$ and $\vec{v}_3=(\sqrt{3/8})(1,1,2/\sqrt{6})^t$. That one of its eigenvalues vanishes, will be crucial when we will discuss the strong coupling limit $U \to \infty$ below.

Using the  eigenvectors $\{\vec{v}_r\}$ it follows
$\vec{\alpha}=(\alpha_1,\alpha_2,\alpha_3)^t= \sum^3_{r=1} \bar{\alpha}_r  \vec{v}_r$
 with $\sum^3_{r=1} \bar{\alpha}_r^2=1$, where  $\bar{a}_r=(\vec{\alpha} \cdot \vec{v}_r)$ are real. Then it is

\begin{equation} \label{eqC9}
\sum^3_{r,r'=1} V_{rr'}\alpha_r\alpha_{r'}=U(\bar{\alpha}_1^2 + 2\bar{\alpha}_2^2) \ .
\end{equation}
The  second line  of Eq.~(\ref{eqC4}) (the constraint for the independent NON, $n_2$) becomes

\begin{equation} \label{eqC10}
1 - \bar{a}_1[\bar{\alpha}_2 + \sqrt{3}\sqrt{1-(\bar{\alpha}_1^2+\bar{\alpha}_2^2)}]=2n_2  \
\end{equation}
from which it follows

\begin{equation} \label{eqC11}
\bar{\alpha}^{(\pm )}_2(n_2;\bar{\alpha}_1)=\Big[(\frac{1}{2}-n_2) \pm \sqrt{3}\sqrt{\bar{\alpha}_1^2(1-\bar{\alpha}_1^2) - (\frac{1}{2}-n_2)^2}\Big]/(2\bar{\alpha}_1) \ .
\end{equation}

$\bar{\alpha}^{(\pm )}_2(n_2;\bar{\alpha}_1)$ put into the r.h.s. of  Eq.~(\ref{eqC9}) yields the functional

\begin{equation} \label{eqC12}
\tilde{\mathcal{F}}[{\bbf{n}};\bar{a}_1]=U[\bar{\alpha}_1^2 + 2\bar{\alpha}^{(\pm )}_2(n_2;\bar{\alpha}_1)^2]  \ ,
\end{equation}
 which has to be minimized with respect to $\bar{a}_1$.
Note that $\bar{\alpha}^{(+)}_2(n_2;\bar{\alpha}_1)  \equiv - \bar{\alpha}^{(-)}_2(1-n_2;\bar{\alpha}_1)$.
 Since $\tilde{\mathcal{F}}[{\bbf{n}};\bar{\alpha}_1]$ involves $\big(\bar{\alpha}^{(\pm )}_2(n_2;\bar{\alpha}_1\big)^2$ the minimization with
 respect to $\bar{\alpha}_1$ yields a functional $\mathcal{F}[n_2]$ exhibiting the particle-hole symmetry
 $\mathcal{F}[n_2]=\mathcal{F}[1-n_2]$ [55].
The resulting equation from that minimization is a polynomial in $\bar{\alpha}_1^2$ of degree \textit{six}. Its roots can not be calculated analytically.   Therefore, we use this situation  to demonstrate the power of a perturbative approach leading in the  weak and strong coupling limit $0 \leq U \ll 1$ and $U \gg 1$, respectively, to asymtotically exact results, obtained analytically.

 \subsection{weak coupling limit}

 The ground state for $U \to 0^{+}$ is twofold degenerate (both states in the first line of Eq.~(\ref{eqC1})).
 In first order in $U$ the degeneracy is lifted and the ground state is given by the state in the first line of Eq.~(\ref{eqC2}). Consequently the coefficients $\alpha_r$ in $|\Psi\rangle$ must fulfil $\alpha_1 \to 1$
 and $\alpha_r \to 0$ for $r=2,3$ which implies $\bar{\alpha}_1 = 1/\sqrt{2} + x_1$and  $\bar{\alpha}_2 = 1/(2\sqrt{2}) + x_2$ with
 $|x_r| \to 0 $.  Taking this small-U dependence into account it follows from Eq.~(\ref{eqC4}) that $n_2 \to 0$,
 i.e., $n_2$ is the \textit{smallness parameter} for the pertubative calculation of $\mathcal{F}$ for weak coupling.  This is intuitively clear, because it is the occupation number of the highest one-particle level $|\nu=2\rangle$. Eq.~(\ref{eqC12}) becomes

\begin{equation} \label{eqC13}
\tilde{\mathcal{F}}[{\bbf{n}};\bar{\alpha}_1]=3U/4 + U [\sqrt{2}(x_1+x_2(n_2;x_1)) + h.o.t.] \ ,
\end{equation}

where $h.o.t.$ stands for higher order terms.
Expanding the r.h.s. of Eq.~(\ref{eqC11}) with respect to $x_1$ leads to

\begin{eqnarray} \label{eqC14}
(x_1+x_2(n_2;x_1))&=&[x_1 - \sqrt{6}\sqrt{n_2 -2x_1^2} \nonumber\\
&+& h.o.t.]/(2(1+\sqrt{2}x_1)) \ ,
\end{eqnarray}
where the minus-sign in Eq.~(\ref{eqC11}) has to be used. The plus-sign has to be chosen for $n_2 \to 1$.
Putting this result into Eq.~(\ref{eqC13}) and minimizing with respect to $x_1$ one obtains

\begin{equation} \label{eqC15}
x^{(min)}_1 = -\sqrt{n_2}/\sqrt{26} +  \mathcal{O}(n_2)  \ .
\end{equation}
The fact that the leading order of $x^{(min)}_1$  is proportional to $\sqrt{n_2}$ justifies a postiori that we have not taken into account the higher order terms in Eq.~(\ref{eqC13}).
Finally substituting $x^{(min)}_1$ into the r.h.s. of Eq.~(\ref{eqC14}) we get from Eq.~(\ref{eqC13})  the functional
in the weak coupling limit

\begin{equation} \label{eqC16}
\mathcal{F}[{\bbf{n}}] = U[3/4 - \sqrt{13/4} \sqrt{n_2} + \mathcal{O}(n_2)] \ .
\end{equation}

\subsection{strong coupling limit}

It is known that the ground state energy of the Hubbard model at half filling converges to zero for $U \to \infty$ \cite{F91}. Therefore, it follows from Eq.~(\ref{eqC9}) that  in the strong coupling limit it must be $\bar{a}_1 \to 0$ and  $\bar{a}_2 \to 0$, i.e. only the eigenvector $\vec{v}_3$ of $(V_{r'r})$ with eigenvalue $0$ contributes on the l.h.s. of Eq.~(\ref{eqC9}) . From Eq.~(\ref{eqC10})  we obtain $n_2 \to (1/2)^{-}$, i.e., $\delta=(\frac{1}{2}-n_2) \geq 0$ is the \textit{smallness parameter} for the perturbative construction of the functional  $\mathcal{F}$ in the strong coupling limit. Since $\bar{\alpha}_1 \to 0$,  Eq.~(\ref{eqC11}) implies first that we have to choose the minus-sign, and second the square root must converge to $\delta$, in order that $\bar{\alpha}_2 \to 0$. The latter condition becomes satisfied if

\begin{equation} \label{eqC17}
\bar{\alpha}_1=2 \delta(1+y_1)/\sqrt{3}  \ .
\end{equation}
Substituting this expression into  Eq.~(\ref{eqC11}) (with the minus-sign) and expanding its r.h.s. with respect to $\delta$ and $y_1$
leads to

\begin{equation} \label{eqC18}
\bar{\alpha}_2(y_1;\delta)=\sqrt{3}\big(\frac{2}{3}\delta^2 - y_1 + h.o.t. \big) \ .
\end{equation}
Next, $\bar{\alpha}_1$ and $\bar{\alpha}_2$ from Eqs.~(\ref{eqC17}) and (\ref{eqC18}) are put into Eq.~(\ref{eqC12}) which yields

\begin{equation} \label{eqC19}
\tilde{\mathcal{F}}[{\bbf{n}},\bar{\alpha}_1] \equiv \tilde{\mathcal{F}}[\delta,y_1]=U\Big[\frac{4}{3}\delta^2(1+y_1)^2 + \frac{1}{3}(2\delta^2 - 3y_1)^2 + h.o.t.\Big]  \ .
\end{equation}
Its minimum is taken at

\begin{equation} \label{eqC20}
y_1(\delta)= \frac{4}{9}\big(\delta^2 - \frac{2}{9}\delta^4 + h.o.t.\big)                  \ .
\end{equation}
In a final step $y_1(\delta)$ is substituted into the r.h.s. of Eq.~(\ref{eqC19}) leading to the \textit{exact} functional

\begin{equation} \label{eqC21}
\mathcal{F}[{\bbf{n}}]=U\Big[\frac{4}{3}(\frac{1}{2}-n_2)^2  + \frac{40}{27}(\frac{1}{2}-n_2)^4 + \mathcal{O}((\frac{1}{2}-n_2)^6)\Big]  \ .
\end{equation}

\end{document}